\documentstyle[12pt]{article}
\newcommand{\be}{\begin{equation}}\newcommand{\ee}{\end{equation}}
\newcommand{\bea}{\begin{eqnarray}}\newcommand{\eea}{\end{eqnarray}}
\newcommand{\ba}{\begin{array}{l}}\newcommand{\ea}{\end{array}}
\begin{document}

\begin{titlepage}
\nopagebreak
\begin{flushright}
UT-679 \\
June 1994\\ hep-th/xxxxxxx \\ ~
\end{flushright}
\begin{center}
{\LARGE
4-dimensional dilaton black holes \\
with cosmological constant}
\vfill
{\large Tadashi Okai}\\
\vspace{1cm}
\it
Department of Physics, University of Tokyo\\
Bunkyo-ku, Tokyo 113, Japan \\
\rm
e-mail: okai@tkyux.phys.s.u-tokyo.ac.jp \\
{}~~~~~~~~~~okai@tkyvax.phys.s.u-tokyo.ac.jp \\
{}~~~~~~~~~~~okai@danjuro.phys.s.u-tokyo.ac.jp
\\
\end{center}
\vfill
\begin{abstract}
Static and spherically symmetric black hole solutions
with non-zero cosmological constant are investigated.
A formal power series solution is found.
It is proved that the number of regular horizons is less than or
equal to 2 for positive cosmological constant and is less than or
equal to 1 for negative cosmological constant.
This shows a striking contrast to the fact that the
Reissner-Nordstr{\o}m-de Sitter black hole with positive cosmological
horizon has 3 regular horizons.
\end{abstract}
\vfill
\end{titlepage}
%
\renewcommand{\theequation}{\thesection.\arabic{equation}}
\setcounter{equation}{0}
\section{Introduction}
There are many models that describe gravity.
And the one, which is especially interesting and we will consider
in this paper, is the model inspired by the string theory\cite{gsw}.
This model is characteristic in that gravity is coupled to the
dilaton field\cite{string}.
Many exact solutions that describe stringy black holes were
found
and their physical properties are discussed
recently\cite{maeda}\cite{garf}\cite{ichinose}.
These solutions
naturally include Einstein's vacuum solutions (Schwarzschild solution,
Kerr solution, and so on), but not solutions of Einstein-Maxwell
theory.
This is because the dilaton $\phi$
has a linear
coupling to $F^2$ ($F_{\mu\nu} = \partial_{\mu}A_{\nu}-\partial_{\nu}A_{\mu}$
is the
field strength of the $U(1)-$gauge field $A_{\mu}$) with
the interaction term of $e^{-2\phi}F^{2}$ in the Lagrangean.
Thus every solution with
nonzero $A_{\mu}$ automatically has non-constant configuration of the
dilaton. It has been revealed that the inclusion of the dilaton
changes the thermal properties\cite{limit}\cite{globalstruct}
and the behaviour
near the extremality of the black holes\cite{elem-part}.
In this paper, we are tempted to examine whether the dilaton changes
physical properties or not when the cosmological term is added to
the action. And if the dilaton does change physical properties,
we want to see how it changes.

We will consider the following action with $a=2$:
\bea
S=\int\sqrt{-g}(R-2(\partial \phi)^{2}-e^{-2a \phi}F^{2}-2\Lambda),
\label{action}
\eea
where $a$ is a constant which determines the strength of the coupling
with the dilaton $\phi$ and the gauge field $A_\mu$.
This action describes the Einstein-Maxwell system for $a=0$, and
Einstein-Maxwell-dilaton system for $a=2$.

Among many physical aspects of black holes, we especially pay
attention to the number of regular horizons.
We already know that
the static and spherically symmetric dilaton
black hole solution (\ref{dilsol}) without cosmological constant
has only one regular horizon and non-zero $F_{\mu\nu}$\cite{garf},
while the Reissner-Nordstr{\o}m solution has 2 regular horizons and
non-zero $F_{\mu\nu}$.
In other words, the global structure of the spacetime with the metric
(\ref{dilsol}) is similar to that of the Schwarzschild solution
rather than the Reissner-Nordstr{\o}m solution.
Based on this example, we are naturally tempted to examine if
such a difference
between the Einstein-Maxwell and the Einstein-Maxwell-dilaton system
can be found or not, when
we move to the case with non-zero cosmological constant.
In other words, our question is that \lq\lq is the global structure of
dilaton black holes with the non-zero cosmological constant
similar to that of the
Schwarzschild-de Sitter solution rather than the
Reissner-Nordstr{\o}m-de Sitter solution?"
The theorem given in the present paper strongly supports
the affirmative answer to this question.
Asymptotically de-Sitter black holes with 3 regular horizons like
the Reissner-Nordstr{\o}m-de Sitter solution are ruled out
in our model.

The organisation of this paper is as follows. In section 2, we will
derive conditions of having regular horizons.
The conditions are concisely expressed in (\ref{assump}). Since we
are considering static and spherically symmetric spacetimes, horizons
and ergosurfaces coincide and they are characterised by $g_{tt}=0$.
When we want to judge whether they are {\it regular} horizons or not,
we need more careful arguments, to which section 2 is mainly devoted.

In section 3, we construct formal power series solutions with
one or more regular horizons. These solutions naturally contain the
Schwarzschild-de Sitter solution as a special case. In this case,
the power series reduces to a polynomial of finite degree.
We show that solutions are either the Schwarzschild-de Sitter or
a formal solution with infinite power series.

In section 4, we consider asymptotically de Sitter spacetimes, in
addition to staticity and spherical symmetry. We prove a theorem
that gives us information on the number of regular horizons.
Making use of the conditions of regular horizons
given in section 2, we will prove that
the number of regular horizons is less than or
equal to 2 for positive cosmological constant and is less than or
equal to 1 for negative cosmological constant.
The theorem shows a striking contrast to the fact that the
Reissner-Nordstr{\o}m-de Sitter black hole with positive (negative)
cosmological horizon has 3 (2) regular horizons.

Finally, we briefly discuss our results in section 5.
%
\setcounter{equation}{0}
\section{Conditions of regular horizons}
We seek static, spherically symmetric solution of the equations of
motion (\ref{eq_motion}). The assumption of staticity and
spherical symmetry reduces the
equations of motion to a suit of ordinary differential equations
that contain several unknown functions of one variable of radial
coordinate $r$\cite{spherical}.
We will see in this section that solving the equations of motion
reduces to solving an integro-differential equation (\ref{g(r)}).
Combining this equation with the condition of the regularity of
horizons, we will get to the conditions of (\ref{assump}).
They are crucial in proving the theorem that gives us the upper limit
of the number of regular horizons, which will be developed in
section 4.
\subsection{Reduction of equations of motion}
The classical equations of motion of (\ref{action}) with coupling
constant $a=2$ are given by:
\bea
&&R_{\mu \nu}=2\partial_{\mu}\phi \partial_{\nu}\phi
+2e^{-2 \phi}F_{\mu \alpha}F_{\nu}^{~~\alpha}
-\frac{1}{2}g_{\mu \nu}e^{-2\phi}F^{2} +g_{\mu\nu}\Lambda,
\label{eq_motion}
\\
&&\Box \phi+\frac{1}{2}e^{-2\phi}F^{2}=0,
{}~~~~\nabla_{\mu}(e^{-2\phi}F^{\mu \nu})
=\frac{1}{\sqrt{-g}}\partial_{\mu}(\sqrt{-g} e^{-2\phi}F^{\mu\nu})=0.
\nonumber
\eea
We can assume the form of the metric, without loss of generality,
to be in the following form\cite{spherical,garf}:
\bea
ds^{2}=-P(r)dt^{2}+\frac{dr^{2}}{P(r)}+Q(r)d\Omega. \nonumber
\eea
The functions $P(r)$ and $Q(r)$ are unknown functions to be determined
by solving the equations of motion.
We pick a Vierbein
\bea
e^{0}=\sqrt{P(r)}dt,~~~~~~~~e^{1}=\frac{dr}{\sqrt{P(r)}},
{}~~~~~~~~e^{2}=\sqrt{Q(r)}d\theta,~~~~~~~~
e^{3}=\sqrt{Q(r)}\sin \theta d\varphi. \nonumber
\eea
The tetrad components of the Ricci tensor $R_{ij}e^{i}\otimes e^{j}$
are calculated as:
\bea
R_{00}&=&\frac{P''}{2}+\frac{P'Q'}{2Q},~~~~~
R_{11}=-\frac{P''}{2}-\frac{P'Q'}{2Q}
+\frac{PQ'^{2}}{2Q^{2}}-\frac{PQ''}{Q}, \nonumber\\
R_{22}&=&R_{33}=
\frac{1}{Q}-\frac{P'Q'}{2Q}-\frac{PQ''}{2Q}.
\eea
And we seek purely magnetic solutions here. We put the form of the
electromagnetic field as
\bea
F_{\mu\nu}dx^{\mu}\wedge dx^{\nu}=h(r)e^{2}\wedge e^{3}
=q \sin \theta d\theta\wedge d\varphi, \nonumber
\eea
where $h(r)$ is again an unknown function and $q$ denotes the magnetic
charge.
Introducing a new function $Z(r):=P(r)e^{2\phi}$,
the equations of motion are equivalent to
\bea
(PQ)''=-4\Lambda Q+2,~~~~~
\left( \frac{PQZ'}{Z}\right)'=-2\Lambda  Q,~~~~~
\phi '^{2}-\frac{Q'^{2}}{4Q^{2}}+\frac{Q''}{2Q}=0.
\nonumber
\eea
The 1st and 2nd equations are integrated with integration constants
$(a,b,c,\phi_{0})$ as:
\bea
PQ=-4\Lambda \int \!\!\! \int Q+r^{2}+ar+b,~~~~~
\log Z=\int \frac{-2\Lambda \int Q+c}
{-4\Lambda \int \!\!\! \int Q+r^{2}+ar+b}-2\phi_{0}.
\label{int}
\eea
Here we should note that we denoted the integration
$\displaystyle \int Q$ and
$\displaystyle \int \!\!\! \int Q$
by {\lq\lq}the primitive function of
$\displaystyle Q(r)$ and
$\displaystyle \left( \int Q\right)\left( r\right)$," respectively.
\it
We explicitly write integration constants {\rm $(a,b,c,\phi_{0})$}
in order to make clear the proof of the theorem in section 4.
\rm
Putting the expression of $(PQ)(r)$ and $\log Z(r)$ in the 3rd
equation of motion, we obtain an integro-differential equation of
the only one unknown function $Q(r)$:
\bea
g(r):=Q(r)A^{2}(r)+2Q'(r)A(r)B(r)+2Q''(r)B^{2}(r)=0,
\label{g(r)}
\eea
where we have introduced two functions which are determined by $Q(r)$;
\bea
&&A(r):=2\Lambda \int Q(r) -2r-a+c, \nonumber\\
&&B(r):=-4\Lambda \int \!\!\! \int Q(r) +r^{2}+ar+b. \nonumber
\eea
Thus we have seen that all the equations of motion are equivalent to
the ordinary integro-differential equation (\ref{g(r)}).
The rest of all the unknown functions are in principle obtained by
quadratures if we can get the solutions of (\ref{g(r)}).
\subsection{Regularity condition of horizons}
We start seeking solutions of eq.(\ref{g(r)}) with regular
horizons. Let us recall that the horizon is defined as points in
which the infinite redshift occurs.
As we are treating only spherical solutions,
these points are clearly defined by
the zeros of the metric component $g_{tt}$, that is, zeros of $B(r)$.
Here we should carefully check if zeros of $B(r)$ contain divergence
of curvature or not. Let us consider the following two situations
in order to determine whether the elements of
$\{ r | B(r)=0  \}$ are regular horizons or curvature singularities.

{\it First situation.}
\\
Suppose $B(r)$ becomes $0$ at a certain point $p$: $B(p)=0$,
then eq.(\ref{g(r)}) demands $Q(r)\sim 0$ around $r=p$ provided
$A(p)\neq 0$. Remembering that the curvature diverges when
$Q(r)=0$, \lq\lq $A(p)\neq 0$ and $B(p)=0$" means that $r=0$ is
a singularity.

{\it Second situation.}
\\
Suppose this time that both $B(p)=0$
and $A(p)=0$ hold at a certain point $r=p$. And suppose further,
for simplicity, that the order of zeros of $B(p)=0$ and $A(p)=0$
are both $1$. Then we can put $k:= B(p)/A(p)$,
which takes some value of neither $0$ nor $\infty$, since we have
supposed that the order of zeros of $B(p)=0$ and $A(p)=0$ are the
same.
And the behaviour of the eq.(\ref{g(r)}) around $r\sim p$ will be
\bea
0=g(r)\sim Q(r)+2k Q'(r)+2k^2 Q''(r).
\nonumber
\eea
Clearly, this admits formal power series solutions of $Q(r)$
around $r=p$.
One can obviously consider more general situations, such as the
order of zeros of $B(p)=0$ and $A(p)=0$ is different. We will not,
however, treat such exceptional situations later on.
\\

Now, we already know two exact solutions of (\ref{g(r)}). They are
the Schwarzschild-de Sitter solution and the dilaton black hole
solution with $\Lambda =0$. Let us look into these two solutions to
make the argument written above more transparent. First, the metric
of the Schwarzschild-de Sitter spacetime is given by\cite{cosmo}
\bea
ds^{2}=-\left( 1-\frac{2M}{r}-\frac{\Lambda}{3}r^{2} \right) dt^{2}
+dr^{2}/\left( 1-\frac{2M}{r}-\frac{\Lambda}{3}r^{2} \right)
+r^{2}
(d\theta^{2}+\sin^{2}\theta d\varphi^{2}).
\nonumber
\eea
One can reproduce this solution by putting $Q(r)=r^2$, $c=-a=2M$,
and $b=0$ in our notation. $M$ is a free parameter. Then $A(r)$ and
$B(r)$ become
\bea
A(r)&=&\frac{2}{3}\Lambda r^{3}-2r-2a,~~~~~
B(r)=-\frac{1}{3}\Lambda r^{4}+r^{2}+ar \label{s-dsAB}
\eea
If $1 \gg \Lambda >0$ and $M>0$, $B(r)$ has 4 zeros of, say,
$r=r_{-} <0$,
$r=0$, $r=r_{+}>0$, and $r=r_{++}>0$. $r=r_{+}$ and $r=r_{++}$
correspond to the black hole horizon and the cosmological horizon,
respectively. If $1 \gg (-\Lambda) >0$ and $M>0$, $B(r)$ has 2 zeros
of $r=0$ and $r=r_{+}>0$. One can easily see that
eq.(\ref{s-dsAB}) leads to
$rA(r)+B(r)=0$ for all $r$. This implies that the zeros and their
multiplicity of $A(r)$ and $B(r)$ are exactly the same, except for
$r=0$. And $r=0$ is a curvature singularity, as can be seen from
$Q(r)=r^2 =0$ at $r=0$. It is an example of the singularity
in the style of the first situation written above. Clearly,
$r=r_{+}>0$ and $r=r_{++}>0$ are regular horizons and they
are examples of the second situation.

As the second example,
we examine the static and spherically symmetric dilaton
black hole solution
with $\Lambda =0$. The metric is given by\cite{garf}
\bea
ds^{2}=-\left( 1-\frac{2M}{r}\right) dt^{2}+dr^{2}
/\left( 1-\frac{2M}{r} \right)
+r(r-\beta)(d\theta^{2}+\sin^{2}\theta d\varphi^{2}).
\label{dilsol}
\eea
This solution is obtained by putting $Q(r)=r(r-\beta)$,
$c=2M-\beta$, and $b=2M\beta$,
where $M$ and $\beta$ are free parameters.
Then $A(r)$ and $B(r)$ become
\bea
A(r)=-2(r-2M), \;\;\; B(r)=(r-\beta) (r-2M).
\nonumber
\eea
Zeros of $B(r)$ are clearly given by $r=2M$ and $r=\beta$. $r=2M$ is
a zero of $A(r)$ as well, and it is a regular horizon. So, this is
an example of the first situation.
$r=\beta$,
on the other hand, is not a zero of $A(r)$. This is in tune with
$Q(r=\beta)=0$, which is a curvature singularity. This is an example
of the second situation.
(Notice that the
dilaton black hole solution is realised by putting the free parameters
as $\beta < 2M$ so that the singularity $r=\beta$ is surrounded by
the regular horizon $r=2M$.)

Summing up the argument of this subsection,
we will seek, from now on, black hole solutions of eq.(\ref{g(r)})
together with the following two assumptions of regular horizons:
\bea
&&\mbox{
        \it
        1) $ \{ \mbox{zeros of $B(r)$}  \} \subseteq
        \{ \mbox{zeros of $A(r)$}  \} $
        \rm ( \it i.e., $B(r)=0 \Longrightarrow A(r)=0$ \rm ),
  }
\nonumber \\
&&\mbox{\it 2) the order of common zeros of $A(r)$ and $B(r)$ is 1.
  }
\label{assump}
\eea
\rm
These conditions will play a crucial role when we prove that the
number of regular horizons is less than or equal to 2, as will be
seen in section 4.
%
\setcounter{equation}{0}
\section{Power series solutions}
In this section, we rewrite the equation of motion (\ref{g(r)}) in
the form of the 4th order ordinary differential equation (\ref{1ode}).
And then we show that eq.(\ref{1ode}) admits formal power series
solutions with a regular horizon.

First, let us change the variable $r$ into $\xi :=r-a/2-c$.
The unknown functions $A(r)$ and $\Lambda Q(r)$ are rewritten as
\bea
A=-\frac{1}{2}\frac{dB}{d\xi}-\xi, \;\;\;
\Lambda  Q=-\frac{1}{4}\frac{d^{2}B}{d^{2}\xi}+\frac{1}{2}. \nonumber
\eea
Plugging these into the eq.(\ref{g(r)}), we obtain
\bea
\left(\frac{d^2 B}{d\xi^2} -2\right)
\left(\frac{1}{2}\frac{dB}{d\xi}+\xi \right)^{2}
-2\frac{d^3 B}{d \xi^3}\left(\frac{1}{2}\frac{dB}{d\xi}+\xi \right)
B+2\frac{d^4 B}{d \xi^4}B^{2}=0.
\label{1ode}
\eea

We will now construct formal power series solutions of (\ref{1ode})
around a regular horizon, say, $\xi =\xi_{+}$. Recalling the
assumptions
of regular horizons (\ref{assump}), we impose conditions on $B(\xi )$
as
\bea
B(\xi _{+})=0, \;\;\; \mbox{and} \;\;\;
\frac{dB}{d\xi }(\xi _{+})= -2\xi _{+}.
\label{regcond}
\eea
Formal power series solutions
$\displaystyle B(\xi )=\sum_{n=0}^{\infty } a_{n} \xi ^{n}$
with a regular horizon $\xi =\xi _{+}$ are constructed.
We prove it
in the style of a proposition:
\\
\\
{\bf proposition }
\it
Formal power series solutions
$\displaystyle B(\xi )=\sum_{n=0}^{\infty } a_{n} \xi ^{n}$
with the condition (\ref{regcond}) is constructed so that each
coefficient $a_{n}$ $(n \geq 0)$ is determined order by order
without any ambiguity. $\displaystyle \frac{~~~~~~}{~~~~~~}$
\rm
\\
\\
{\bf proof } One can easily notice that eq.(\ref{1ode}) is invariant
under the (global) scale transformation
\bea
\xi  \longmapsto \alpha  \xi , \;\;\;
B \longmapsto \alpha ^{2}B \;\;\;
(\;\; \alpha\in {\bf R}\setminus\{ 0 \} \;\; ).
\nonumber
\eea
So, one can assume $\xi _{+}$ be $0$ or $\pm 1$ without loss of
generality. $\xi _{+}=0$ corresponds to an exceptional case we have
mentioned in the last section, and we will put it aside. Since
there is
no substantial difference between $\xi _{+}=+1$ and $\xi _{+}=-1$ when
arguing the construction of formal solutions, we will set
$\xi _{+}=-1$. Setting $\xi +1=:u$ for convenience, solving (\ref{1ode})
is equivalent to finding a solution
$\displaystyle B(u)=\sum_{n=1}^{\infty } a_{n} u^{n}$
that satisfies
\bea
\tilde{g}(u):=
\left( \frac{d^2 B}{du^2}-2\right)
\left( \frac{1}{2}\frac{dB}{du}+u-1 \right)^{2}
-2\frac{d^3 B}{du^3}
\left( \frac{1}{2}\frac{dB}{du}+u-1 \right) B
+2\frac{d^4 B}{du^4}B^{2}=0.
\nonumber
\eea

Notice that $B(\xi _{+})=0$ and
$\displaystyle \frac{dB}{d\xi }(\xi _{+})=-2\xi _{+}$ are fulfilled by
putting $a_{0}=0$ and $a_{1}=-2$, respectively.
We define $\{ b_i \}_{i \geq 1}$, for convenience, as
\bea
b_{1}:=0,\;\;\; b_{2}:=a_{2}+1,\;\;\; b_{i}:=a_{i} \;\;(\; \mbox{for}
\;\; i \geq  3 \;).
\nonumber
\eea
Then, each term in the right hand side of $\tilde{g}(u)$ is
calculated as:
\bea
&&B(u)=2u-u^{2}+\sum_{i=2}^{\infty}b_{i}u^{i}, \;\;\;
\frac{1}{2}\frac{dB}{du}(u)+u-1
        =\frac{1}{2}\sum_{i=2}^{\infty}ib_{i}u^{i-1},
\nonumber  \\
&&\frac{d^2 B}{du^2}(u)-2=-4+\sum_{i=2}^{\infty}i(i-1)b_{i}u^{i-2},
\;\;\;
\frac{d^3 B}{du^3}(u)=\sum_{i=3}^{\infty}i(i-1)(i-2)b_{i}u^{i-3},
\nonumber \\
&&\frac{d^4 B}{du^4}(u)
        =\sum_{i=4}^{\infty}i(i-1)(i-2)(i-3)b_{i}u^{i-4}.
\nonumber
\eea

Putting these into $\tilde{g}(u)$, the coefficient of $u^i$ in
$\tilde{g}(u)$ is given by
\bea
&&-\sum_{(\# 1)}qrb_{q}b_{r}
+\frac{1}{4}\sum_{(\# 2)}p(p-1)qr b_{p}b_{q}b_{r}
\label{ui-coeff} \\
&&-\sum_{(\# 3)}p(p-1)(p-2)q b_{p}b_{q}b_{r}
+\sum_{(\# 4)}p(p-1)(p-2)q b_{p}b_{q}
-\sum_{(\# 5)}2p(p-1)(p-2)q b_{p}b_{q} \nonumber \\
&&+\sum_{(\# 6)}2p(p-1)(p-2)(p-3) b_{p}b_{q}b_{r}
-\sum_{(\# 7)}4p(p-1)(p-2)(p-3) b_{p}b_{q} \nonumber\\
&&+\sum_{(\# 8)}8p(p-1)(p-2)(p-3) b_{p}b_{q} \nonumber \\
&&+2i(i-1)(i-2)(i-3)b_{i}
+8(i+1)i(i-1)(i-2)b_{i+1}
+8(i+2)(i+1)i(i-1)b_{i+2},
\nonumber
\eea
where the conditions of the summation of (\# 1), $\cdots$, (\# 8) are
listed as:
\bea
&&(\# 1):=\{ q\geq  2,\;\; r\geq  2,\;\; q+r=i+2  \}, \nonumber \\
&&(\# 2):=\{ p\geq  2,\;\; q\geq  2,\;\; r\geq  2,\;\; p+q+r=i+4  \},
\nonumber \\
&&(\# 3):=\{ p\geq  3,\;\; q\geq  2,\;\; r\geq  2,\;\; p+q+r=i+4  \},
\nonumber \\
&&(\# 4):=\{ p\geq  3,\;\; q\geq  2,\;\; p+q=i+2  \}, \nonumber \\
&&(\# 5):=\{ p\geq  3,\;\; q\geq  2,\;\; p+q=i+3  \}, \nonumber \\
&&(\# 6):=\{ p\geq  4,\;\; q\geq  2,\;\; r\geq  2,\;\; p+q+r=i+4  \},
\nonumber \\
&&(\# 7):=\{ p\geq  4,\;\; q\geq  2,\;\; p+q=i+2  \}, \nonumber \\
&&(\# 8):=\{ p\geq  4,\;\; q\geq  2,\;\; p+q=i+3  \}.
\nonumber
\eea
After long calculations, one sees that $u^{i}-$coefficient of
(\ref{ui-coeff}) is written in terms of
$\left\{ b_2\right.$, $b_3$, $\cdots$, $\left. b_{i+2}\right\}$
only, and that the
$b_{i+2}-$dependence of (\ref{ui-coeff}) is linear:
\bea
8(i+2)(i+1)i(i-1)b_{i+2}+(\mbox{terms that do not contain $b_{i+2}$}).
\nonumber
\eea
This indicates that $b_3$, $b_4$, $\cdots$ are determined step by
step without ambiguity and expressed in terms of $b_2$ and $b_3$.
{}~~~~~ \rule{4mm}{4mm}
\\

To conclude this section, we show that the power series solution
is either in the form of polynomials of degree 4 or in the form of
infinite sum.
Suppose the solution
$\displaystyle B(u)=\sum_{i=1}^{\infty } a_{i} u^{i}$
be a polynomial of $u$ and let $p$ be its degree. That is,
$a_{p} \neq 0$ and $a_{p+1}=a_{p+2}=\cdots =0$.
Then the term of the highest degree with respect to $u$ in
$\tilde{g}(u)$ is calculated and given by
\bea
&&\left\{ p(p-1)\times  (p/2)^{2}-2p(p-1)(p-2)(p/2)\times  1
+2p(p-1)(p-2)(p-3)\times  1^{2} \right\} a_{p} u^{3p-4}
\nonumber \\
&=&\frac{1}{4}p(p-1)(p-4)(5p-12)a_{p} u^{3p-4}.
\nonumber
\eea
Eq.(\ref{1ode}) demands it should vanish.
Both $p=0$ and $p=1$, however, contradict the finiteness of curvature
$Q(u) \neq 0$ and thus they are meaningless. $p=12/5$ is excluded by
way of the argument that follows. Assumptions of the regularity of
the horizon demand $B(u=0)=0$ and $\displaystyle \frac{dB}{du}(u=0)=2$.
So, we have no choice but to put
\bea
B(u)=2u+\alpha  u^{7/5}+\beta  u^{2}+\gamma  u^{12/5}
\nonumber
\eea
with some real constants $\alpha$, $\beta$, and $\gamma$.
But $\tilde{g}(u)=0$ leads only to $\alpha =\beta =\gamma =0$.
This is a contradiction.

Further calculation reveals that the polynomial solution with its
degree 4 is the well-known Schwarzschild-de Sitter solution. Therefore
the new formal solution is written in the form of an infinite
power series of the radial coordinate $r$. Unfortunately, the radius
of convergence of this formal solution is at present not known.
It is strongly hoped that this formal solution will be written in
an analytically closed form through some nice change of coordinates
or some other ways, and that more detailed properties can be cast
from it.
%
\setcounter{equation}{0}
\section{A theorem on the number of regular horizons}
In this section, we prove the following theorem:
\\ \\
{\bf theorem } \\
\it
1) If $\Lambda >0$, the number of regular horizons of the black hole
solutions which is asymptotically de-Sitter is less than 2. \\
2) If $\Lambda <0$, the number of regular horizons of the black hole
solutions which is asymptotically de-Sitter is less than 1.
$\displaystyle \frac{~~~~~~}{~~~~~~}$
\\
\\
\rm
We provide the two lemmas in the following
to prove this theorem.\\
\\
{\bf lemma 1 }
\it
If $Q(r)$ satisfies the equation (\ref{g(r)}), there does not exist
a point $p$ that satisfies
\rm
\bea
\left\{ Q(p)>0, Q'(p)=0, Q''(p)>0 \right\}. \frac{~~~~~~}{~~~~~~}
\nonumber
\eea
{\bf proof of lemma 1 }
We suppose there exists such a point $p$, and show that it leads us to
contradiction.
First, by assumption of this lemma,
\bea
g(p)=Q(p)A^{2}(p)+0+2Q''(p)B^{2}(p)=0.
\nonumber
\eea
The assumption $Q(p)>0$ and $Q''(p)>0$ require $A(p)=B(p)=0$.
Now, we put for simplicity
\bea
\left( \frac{d^{i}g}{dr^{i}} \right)(r)=:g_{i}(r), \;\;\;
\left( \frac{d^{i}A}{dr^{i}} \right)(r)=:A_{i}(r), \;\;\;
\left( \frac{d^{i}B}{dr^{i}} \right)(r)=:B_{i}(r). \;\;\;
\nonumber
\eea
The equation of motion (\ref{g(r)}) requires $g(r)$ identically vanish
on the whole region. This means that all orders of derivatives of
$g(r)$ be zero.
{}From $Q'(p)=A(p)=B(p)=0$,
$g_{1}(p)=0$ is automatically satisfied.
Now, let us observe $g_{2}(p)$:
\bea
g_{2}(p)=2QA'^{2}+0+4Q''B'^{2}. \nonumber
\eea
Equation of motion requires $g_{2}(p)=0$. Using the assumption of
$Q(p)>0$ and $Q''(p)>0$ again, we have $A_{1}(p)=B_{1}(p)=0$.
We will show by induction that $A_{n}(p)=B_{n}(p)=0$ holds for all
positive integer $n$.
Actually, we already know it it true for $n=0$. We suppose that it is
true for up to n:
\bea
A_{0}(p)=A_{1}(p)=\cdots =A_{n}(p)=0, \;\;\;
B_{0}(p)=B_{1}(p)=\cdots =B_{n}(p)=0.
\nonumber
\eea
Then,
\bea
&&(QA^{2})_{0}(p)=(QA^{2})_{1}(p)=\cdots =(QA^{2})_{2n+1}(p)=0,
\nonumber \\
&&(QA^{2})_{2n+2}(p)
=\left( \matrix{ 2n+2 \cr n+1 \cr } \right)
Q(p)(A_{n+1}(p))^{2} \;\;\; \mbox{hold}.
\nonumber
\eea
Similar arguments for $2(Q'AB)$ and $(Q''B^2)$ lead to
\bea
&&2(Q'AB)_{0}(p)=2(Q'AB)_{1}(p)=\cdots =2(Q'AB)_{2n+2}(p)=0,
\nonumber \\
&&2(Q'AB)_{2n+3}(p)=2(n+2)
\left(  \matrix{ 2n+3 \cr n+2 \cr } \right)
Q''(p) A_{n+1}(p) B_{n+1}(p)
\nonumber
\eea
and
\bea
&&2(Q''B^{2})_{0}(p)=2(Q''B^{2})_{1}(p)
=\cdots =2(Q''B^{2})_{2n+1}(p)=0,
\nonumber \\
&&2(Q''B^{2})_{2n+2}(p)=
2\left(  \matrix{ 2n+2 \cr n+1 \cr } \right)
Q''(p)(B_{n+1}(p))^{2}.
\nonumber
\eea
Therefore, under the assumption of
\bea
&&Q(p)>0,\;\;\; Q'(p)=0, \;\;\; Q''(p)>0, \nonumber \\
&&A_{0}(p)=A_{1}(p)=\cdots =A_{n}(p)=0, \;\;\;
B_{0}(p)=B_{1}(p)=\cdots =B_{n}(p)=0, \nonumber
\eea
we obtain
\bea
&&g_{0}(p)=g_{1}(p)=\cdots =g_{2n+1}(p)=0, \nonumber \\
&&g_{2n+2}(p)=
\left( \matrix{ 2n+2 \cr n+1 \cr } \right)
Q(p)(A_{n+1}(p))^{2}
+2\left(  \matrix{ 2n+2 \cr n+1 \cr } \right)
Q''(p)(B_{n+1}(p))^{2}.
\nonumber
\eea
The equation of motion $g(r)=0$ requires $g_{2n+2}(p)=0$. Thus we
obtain $A_{n+1}(p)=B_{n+1}(p)=0$. Now the proof of the induction is
completed and $A_{n}(p)=B_{n}(p)=0$ holds for all positive integer
$n$.

Clearly, on the other hand, $Q''(p)>0$ and $B_4 (p)$
$=-4\Lambda Q''(p)$ $=0$ contradict each other, and thus
lemma 1 is proved.
{}~~~~~ \rule{4mm}{4mm}
\\
\\
{\bf lemma 2 }
\it
$Q(r)$ never decreases with respect to $r$ as long as $Q(r)>0$ holds.
\rm
$\frac{~~~~~~}{~~~~~~}$
\\
\\
{\bf proof of lemma 2 }
\\
(1st step)
We suppose that there exists a local minimum at $r=p$ and show that
it will lead to a contradiction. That is to say, we suppose that
\bea
Q(p)>0,\;\;\; Q_{1}(p)=Q_{2}(p)=\cdots =Q_{2l-1}(p)=0, \;\;\;
Q_{2l}(p)>0 \nonumber
\eea
hold for some $l \in \{1,2,3, \cdots \}$ and get to a contradiction.
Here we have put
$ \displaystyle Q_{i}:=\frac{d^{i}Q}{dr^{i}}$.
Under this assumption,
the derivatives of $g(r)$ at $r=p$ are calculated as follows.
First, $g_0 (p) = Q(p)(A(p))^2$. From the equation of motion
(\ref{g(r)}), we get $A(p)=0$. Then $g_1 (p) = 2Q(p)A(p)A_1 (p)$ and
$g_1 (p) =0$ automatically holds. Next, $g_2 (p)$ is calculated as
$g_2 (p) = 2Q(p) ( A_1 (p))^2$. From the equation of motion again,
we get $A_1 (p) =0$. Continuing these arguments,
$g_{2l-4}$ is calculated as $\displaystyle
g_{2l-4}(p)$
$\displaystyle =\left( \matrix{2l-4 \cr l-2 \cr}\right)  $
$\displaystyle Q(p)(A_{l-2}(p))^{2}$ and it leads to $A_{l-2}(p)=0$.
Then $g_{2l-3}(p)=0$ automatically holds and $g_{2l-2}(p)$ is
calculated as $\displaystyle g_{2l-2}(p)$
$\displaystyle =\left( \matrix{2l-2 \cr l-1 \cr}\right)$
$\displaystyle Q(p)(A_{l-1}(p))^{2}+2Q_{2l}(p)(B(p))^{2}$
and it leads to $A_{l-1}(p)=B_0 (p)=0$.

Here we rewrite the algorithm of calculations written above:
\bea
&&g(r)=QA^{2}+2Q'AB+2Q''B^{2}=0 \nonumber \\
\mbox{step } 0: &&g_{0}(p)=QA^2 \;\;\; \Longrightarrow \;\;\; A(p)=0.
\nonumber \\
\mbox{step } 1: &&g_{1}(p)=2QAA'=0 \;\;\;
        \mbox{automatically holds.}
\nonumber \\
\mbox{step } 2: &&g_{2}(p)=2QA_{1}^{2} \;\;\; \Longrightarrow
        \;\;\; A_{1}(p)=0.
\nonumber \\
&&\;\;\; \vdots  \;\;\; \;\;\; \;\;\; \;\;\; \vdots
\nonumber \\
\mbox{step } (2l-4): &&g_{2l-4}(p)
        =\left( \matrix{2l-4 \cr l-2 \cr}\right)
        QA_{l-2}^{2} \;\;\;
        \Longrightarrow \;\;\; A_{l-2}(p)=0.
\nonumber \\
\mbox{step } (2l-3): &&g_{2l-3}(p)=0
        \;\;\;\mbox{automatically holds.} \nonumber \\
\mbox{step } (2l-2): &&g_{2l-2}(p)
        =\left( \matrix{2l-2 \cr l-1 \cr}\right)
         QA_{l-1}^{2}+ 2Q_{2l}B^{2}
        \;\;\; \Longrightarrow \;\;\; A_{l-1}(p)=B_{0}(p)=0.
\nonumber
\eea
(2nd step)
Next, we will prove by induction that $A$, $B$, and all the orders of
their derivatives vanish at $r=p$.
We already know that
\bea
&&A_{0}(p)=A_{1}(p)=\cdots =A_{l-1+n}(p)=0, \nonumber \\
&&B_{0}(p)=B_{1}(p)=\cdots =B_{n}(p)=0, \nonumber
\eea
is true for $n=0$.
Now we assume that this is true up to a certain positive integer $n$.
Then, $g_{2l+2n}(p)$ is calculated as
\bea
g_{2l+2n}(p)
        =\left( \matrix{2l+2n \cr l+n \cr}\right)
          Q (A_{l+n})^{2}
        +\left( \matrix{2l+2n \cr 2l-2 \cr}\right)
         \left( \matrix{2n+2 \cr n+1 \cr}\right)
         Q_{2l}(B_{n+1})^{2}.
\nonumber
\eea
Thus, using the assumptions of $Q>0$ and $Q_{2l}>0$, we obtain
\bea
A_{l+n}(p)=B_{n+1}(p)=0. \nonumber
\eea
This means that the proof of the induction is finished.
By the same argument given
in lemma 1, $Q_{2l}(p)>0$ and $B_{2l+2}(p)=0$ contradict each other
and the proof of lemma 2 is now completed.
{}~~~~~ \rule{4mm}{4mm}
\\
\\
{\bf proof of the theorem }
Since we have explicitly written the integral constants
$(a, b, c)$ in (\ref{int}), we can let
\bea
\left(\int Q\right)(r)= \int_{r_{0}}^{r}\!\! dx Q(x), \;\;\;
\left(\int \!\!\! \int Q\right)(r)=
\int_{r_{0}}^{r}\!\! dx \! \int_{r_{0}}^{x}\!\! dy Q(y)
\nonumber
\eea
without loss of generality.
Here we denoted that $r=r_{0}$ is the position of the regular horizon.
\\
1) We will prove the theorem for $\Lambda > 0$. By definition,
we get
\bea
A(r)=2\Lambda \left( \int Q\right)(r) -2r+a+c,\;\;\;
\frac{dA}{dr}(r)=2\Lambda Q(r) -2,\;\;\;
\frac{d^{2}A}{dr^{2}}(r)=2\Lambda Q'(r).
\nonumber
\eea
Therefore $\displaystyle \frac{dA}{dr}(r_{0})=-2$.
{}From lemma 1,
\bea
\frac{d^2 A}{dr^2}(r)\geq 0 \;\;\; \mbox{holds for all}
\;\;\; r\geq r_0.
\label{cond1}
\eea
Since we have assumed that the spacetime is asymptotically de-Sitter,
the behaviour of $Q(r)$ for large $r$ is $Q(r) \sim r^2$.
Therefore
\bea
\left( \frac{dA}{dr} \right) (r=\infty) =\infty.
\label{cond2}
\eea
Using (\ref{cond1}), (\ref{cond2}) and lemma 2, we see that
there exists in the region of $r\geq r_0$ only one local minimum of
$A(r)$ at, say, $r=r_1$. That is to say, the $r_1$
that satisfies $r_1 \geq r_0$ and
$\displaystyle \frac{dA}{dr}(r_1)=0$ is uniquely determined.
Thus, $A(r)$ decreases when $r_0 \leq r \leq r_1$ and increases
when $r_1 \leq r$ as $r$ becomes larger.
This indicates that the number of zeros of $A(r)$, which is the
number of regular horizons, is less than or equal to 2.
This completes the proof for $\Lambda > 0$.
\\
\\
2) If $\Lambda <0$,
\bea
\frac{dA}{dr}(r) = -(-2\Lambda) Q(r)-2 <0
\nonumber
\eea
holds for all $r \geq r_0$. Using lemma 2,
$A(r)$ monotonously decreases
for $r \geq r_0$ as $r$ becomes larger. This indicates the number
of regular horizons is less than or equal to 1.
{}~~~~~ \rule{4mm}{4mm}
%
\setcounter{equation}{0}
\section{Discussion}
We showed in section 3
that the solving the equations of motion of the cosmological
dilaton-Einstein-Maxwell system is reduced to solving the 4th order
ordinary differential equation of $B(r)$. We have seen that it
admits formal power series solutions.
Each coefficient of the power series is recursively and unambiguously
determined. The admissible order of these power series is 4 or
$\infty$. The former corresponds to the well known Schwarzschild-de
Sitter solution and the dilaton field is constant. The
latter corresponds to a new solution in which the dilaton field
is not in
constant configuration. The convergence of the latter solution is,
however, hard to prove and we leave this problem for future
investigation.

In section 4, we proved a theorem which gives the upper bound of
the number of regular horizons. It is interesting that some global
data of the spacetime, such as the number of regular horizons, can be
cast directly from the equations of motion and the boundary
condition ($=$asymptotic condition).
It is rather surprising to the author that the
proof of the theorem is carried out without solving the equations
of motion.
One may naturally wonder
what kinds of global data are known without solving the Einstein
equation and how these data (if they exist) can be obtained.
I hope that the result obtained in section 4 will do some help in
order to answer these questions.
\\
\\
\\
\footnotesize {\it Acknowledgments.}
I would like to thank Professor T.Eguchi for sincere guidance and
suggestion.
Section 3 of the present paper was refined through the discussion
with Professor J.Arafune, to whom I greatly owe for discussion,
continuous encouragement, and valuable comments.
I am grateful to Professor A.Kato and all the members of the
Elementary Particle Theory Group in the University of Tokyo for
support and collaboration.
\normalsize
%
\\


\begin{thebibliography}{99}
\bibitem{spherical} The reader is referred to any textbook of
        general relativity. See, for example, section 6 of
        \cite{wald}, or section 3 of \cite{chandra}.
\end{thebibliography}
\end{document}